\documentclass[a4paper]{article}

\usepackage{icrc2013}
\newcommand{\commentout}[1]{}

\title{Mechanics and cooling system for the camera of the Large Size Telescopes of the Cherenkov Telescope Array (CTA).}

\shorttitle{}

\authors{
Carlos Delgado$^{1}$,
Oscar Blanch$^{2}$,
Carlos Diaz$^{1}$,
Noemi Hamer$^{1}$,
Ohoka Hideyuki$^{3}$, 
Razmik Mirzoyan$^{4}$,
Masahiro Teshima$^{4}$$^{5}$,
Holger Wetteskind$^{4}$,
Tokonatsu Yamamoto$^{6}$
for the Cherenkov Telescope Array Consortium.
}

\afiliations{
$^1$ CIEMAT, Madrid, Spain \\
$^2$ IFAE, Barcelona, Spain \\
$^3$ ICRR, University of Tokyo, Japan \\
$^4$ Max-Planck-Institute for Physics, Munich, Germany \\ 
$^5$ Institute for Cosmic Ray Research, University of Tokyo, Japan \\
$^6$ Konan University, Japan
}

\email{carlos.delgado@ciemat.es}

\abstract{Mechanics of the camera for the large size telescopes of CTA must
protect and provide a stable environment for its instrumentation. This is achieved
by a stiff support structure enclosed in an air and water tight volume. The structure
is specially devised to facilitate extracting the power dissipated by the focal plane
electronics while keeping its weight small enough to guarantee an optimum load on the
telescope structure. Details of this system are shown.}


\begin{document}
\maketitle

\section{Introduction}
The Cherenkov Telescope Array (CTA), described in detail in \cite{bib:report},
will have a tenfold higher sensitivity compared to existing 
ground-based Cherenkov telescope installations and will cover an energy 
range between 20 GeV and 100 TeV or beyond. To cover this wide energy range, CTA will comprise
3 sizes of telescopes: 23 m mirror diameter Large Size Telescopes (LSTs), 10-12 m Mid-Size
Telescopes (MSTs) and 4-6 m Small-Size Telescopes (SSTs). The design of the LST telescopes is partly based on the  
MAGIC telescopes currently operated at La Palma (Spain). One of the 
conceptual differences between the LST camera design and the MAGIC camera one 
is that the readout electronics are hosted within the camera body for the former, 
whereas they are kept in a building close to the telescope for the latter. This, together
with the requirement of operating the electronics under optimum conditions for 
good performance and durability, requires a careful design of
the mechanics of the camera and its interior environmental control system. The
current design of these mechanics are described below, together with studies carried out to verify
their performance.

\section{LST Camera mechanics requirements}
The camera mechanical structure requirements are defined by the environmental
conditions at the chosen CTA site, the specifications on 
the required environmental conditions inside the camera, and on the 
stiffness requirements to guarantee stable position of camera pixels
with respect the telescope optical axis. Additional requirements due to 
interfaces with other systems in the telescope and easy
maintainability of the whole camera system are also taken into account. 

The main requirements are the following:  
\begin{itemize}
\item The camera mechanics is exposed to outside environment, and must provide protection to all 
its components under the survival conditions defined for the CTA observatory. This includes a range of the air temperature from -20$^\circ$C to 40$^\circ$C, with temperature shocks of up to +/- 30$^\circ$C in 24 hours. Additionally, the relative humidity on the site can range from 3\% to 100\% and the air pressure from 600 hPa to 900 hPa. Finally the camera should withstand a sunlight exposure of up to 1200 W/m$^2$ without damaging its components. 
\item The camera should provide a suitable environment for the hosted photo detectors and electronics. This requires that the camera mechanics avoids the entrance of dust and water to its interior, thus should be air tight. To this end the camera should provide an air tight optical window on its front, coplanar to the focal plane. Upon deformation, no part of the entrance window must get in contact with components in the interior of the camera.
\item A blind or similar mechanism should be present to cover the photo-detector plane during standby. It should be able to open and close in any telescope orientation.  
\item To guarantee the best physics performance, the position of the photo detector entrance windows must not be displaced by more than 1 mm with respect to its nominal position during observations. Additionally, the dead space between photo-detectors must be at most 1 mm.  
\item The surface temperature of all electronics hosted in the camera has to be kept constant in time during observations. Additionally, the temperature should be low enough to guarantee a high mean time between failures of all electronics.
\item Further constraints defined by the interface to the camera frame which fixes the camera to the arch, and by the maintainability requirements are that total camera weight, including mechanics and electronics, shall not exceed 2000 kg to avoid stress damages to the telescope structure. Additionally, the camera shall allow easy access to all its components for replacement and a simple procedure to load and unload it for maintenance.
\end{itemize}

\section{Camera mechanics overview}
\begin{figure}[t]
  \centering
	\includegraphics[width=0.35\textwidth]{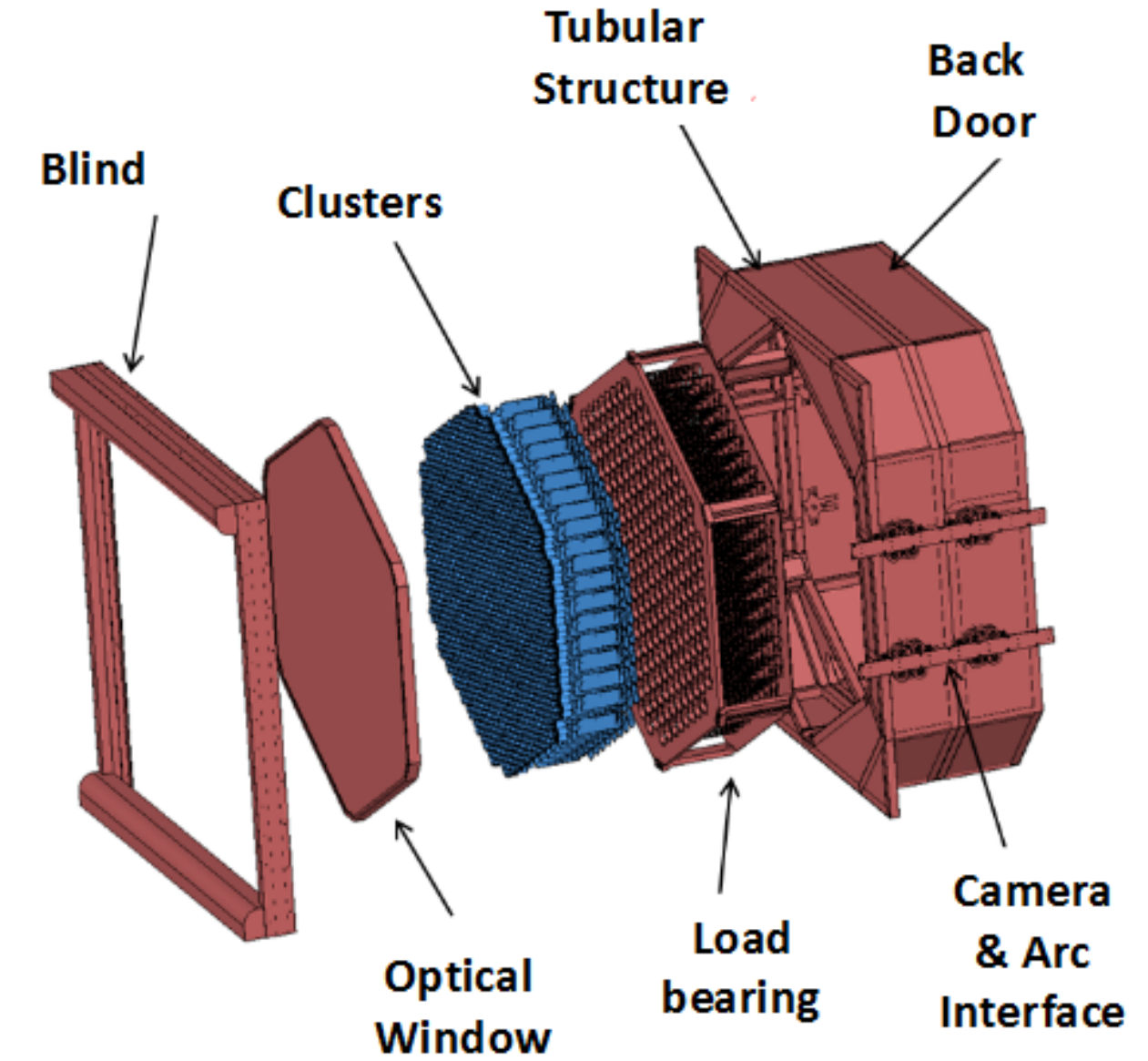}
  \caption{Exploded view of the camera mechanical elements.}
  \label{fig:exploded view}
\end{figure}

The camera mechanics elements are shown in the exploded view of Fig. \ref{fig:exploded view}.
Its dimensions are $2.9\times 2.9\times 1$ m$^3$. Internally, the camera volume is divided 
in two parts. The front one hosts the readout electronics and photodetectors, whereas the
rear one hosts the back plane, cabling, and auxiliary electronics. The instrumented area lies behind a protective
entrance window in the front part of the camera, which also ensures the air tightness of 
the camera volume. The shape of the instrumented area is hexagonal, with a maximum diameter 
of $2.25$ m. Behind the protective surface, light guides coupled to the photo-detectors define
the pixels of the camera. Sets of seven of these photo-detectors are mechanically and 
electrically coupled with the readout electronics, in an ensemble denominated a cluster. These 
clusters are inserted into the load bearing structure, which guarantees the mechanical 
stability of the pixel position, and provides the necessary infrastructure for the cooling
of the electronics. The back side of the load bearing structure has slits that allow to connect the
cluster readout electronics to the active backplanes used for trigger and clock signals
distribution. The load bearing structure is mounted inside 
an outer tubular structure, which provides the necessary surfaces to hold the camera's 
external walls, internal access doors and interfaces with the camera frame in the telescope arch.

\section{Tubular structure}

\begin{figure}[t]
  \centering
	\includegraphics[width=0.2\textwidth]{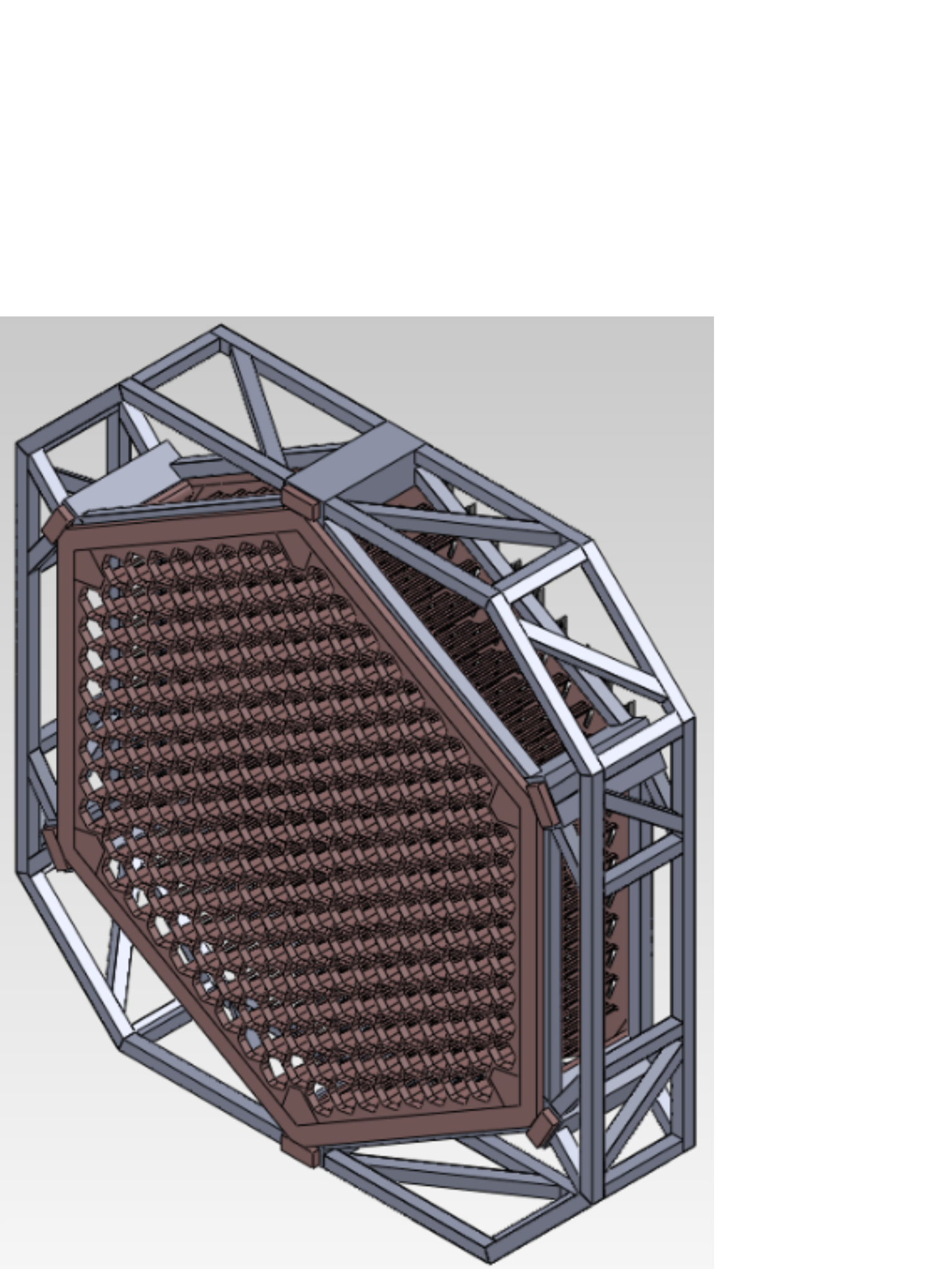}
  \caption{Tubular structure (light colors) and clusters load bearing structure (dark colors).}
  \label{fig:tubular and bearing}
\end{figure}

The main mechanical structural element of the camera is the tubular structure, which provides the support
for all mechanical elements in the camera and the fixation points with the system joining the camera with the telescope arch. This structure, shown with light colors in Fig. \ref{fig:tubular and bearing}, is made of hollow aluminum tubes to keep its weight as low as possible. Finite element analysis of the stiffness shows that under a load of 2500 kg in addition to its weight, the structure shows a maximum deformation in the direction parallel to the optical axis of 2.5 mm, and below 1 mm in any directoin perpendicular to it.

\section{Load bearing structure}
\begin{figure}[t]
  \centering
	\includegraphics[width=0.3\textwidth]{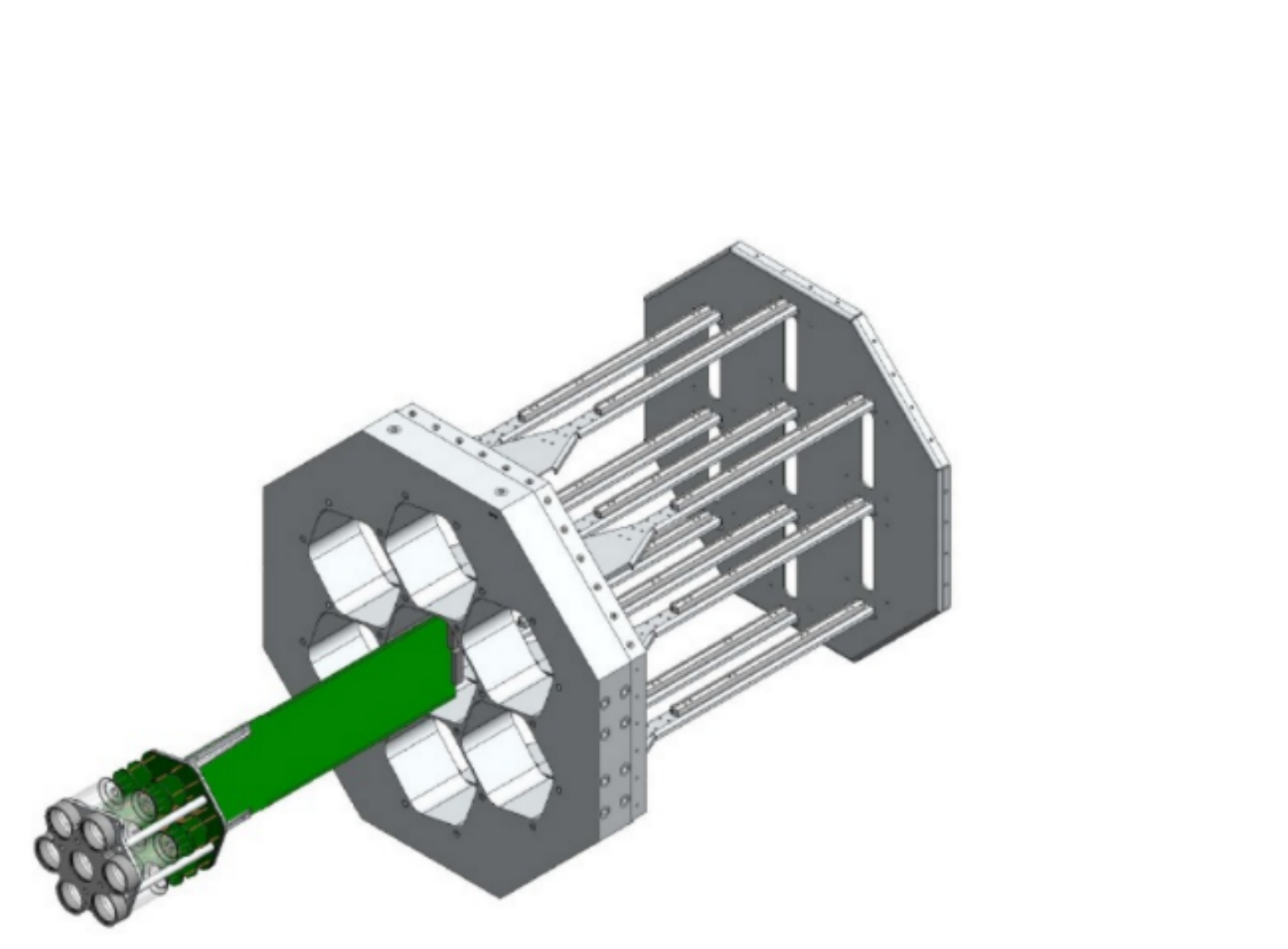}
  \caption{Drawing showing the concept of insertion of a cluster from the front side of the camera in a model of the load bearing structure.} 
  \label{fig:insertion}
\end{figure}

The role of the load bearing structure, shown with dark colors in Fig. \ref{fig:tubular and bearing}, is to hold the photo-detectors and the readout electronics and to provide the necessary infrastructure to keep them in an optimum environment. It consists of a sandwich made of two aluminum plates joined by towers screwed to the plates. The plates have slits where the clusters are inserted from the front of the camera and connected to the back-planes. The joining towers serve as rails to hold the readout boards and guide them during their insertion, as shown in the conceptual drawing of Fig. \ref{fig:insertion}. This structure is inserted into the tubular structure and screwed to it in several locations on the front and rear planes.

Two different designs of the load bearing structure sharing the same interface with the tubular structure have been considered. These provide the necessary infrastructure to control the camera interior temperature 
by means of a water cooling system or an air cooling system, respectively.

\begin{figure}[t]
  \centering
	\includegraphics[width=0.3\textwidth]{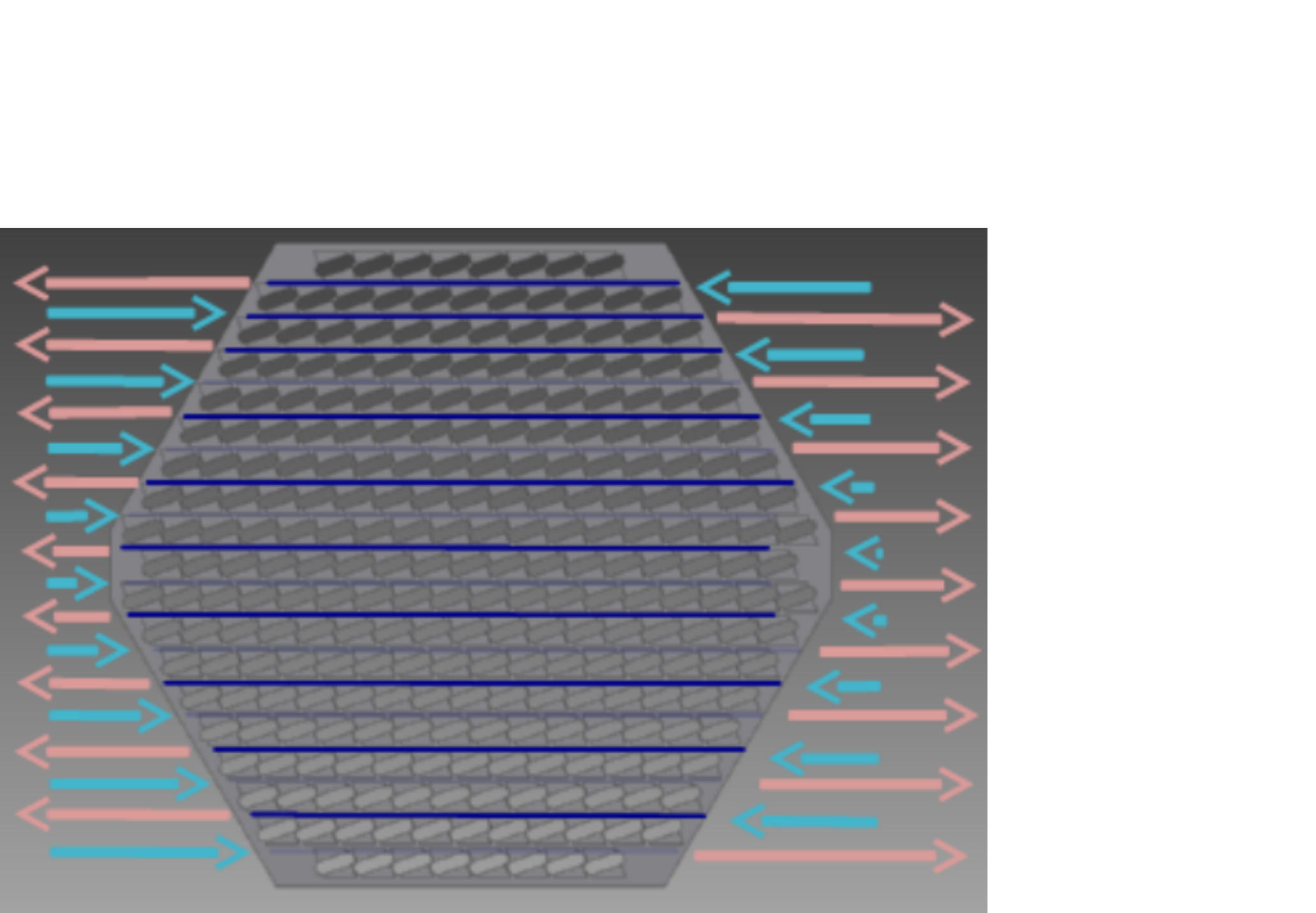}
  \caption{Conceptual drawing showing the position of the channels 
  in the plates of the load bearing structure for the water cooling system (blue arrows), and the return of heated water (red arrows)} 
  \label{fig:water circulation}
\end{figure}

In the case of the structure designed for the water cooling system,
the two 20 mm plates of the load bearing structure have longitudinal channels drilled in their interior. The plates temperature is kept constant by circulating temperature controlled water through these channels, as it is shown in Fig. \ref{fig:water circulation}. By providing an auxiliary mechanics to the cluster electronics (described in section \ref{sec:cluster}), thermal contact between the readout electronics and the plates is obtained. This
allows to keep the temperature of the electronics constant in time and in an optimum range. 
 
 \begin{figure}[t]
  \centering
	\includegraphics[width=0.25\textwidth]{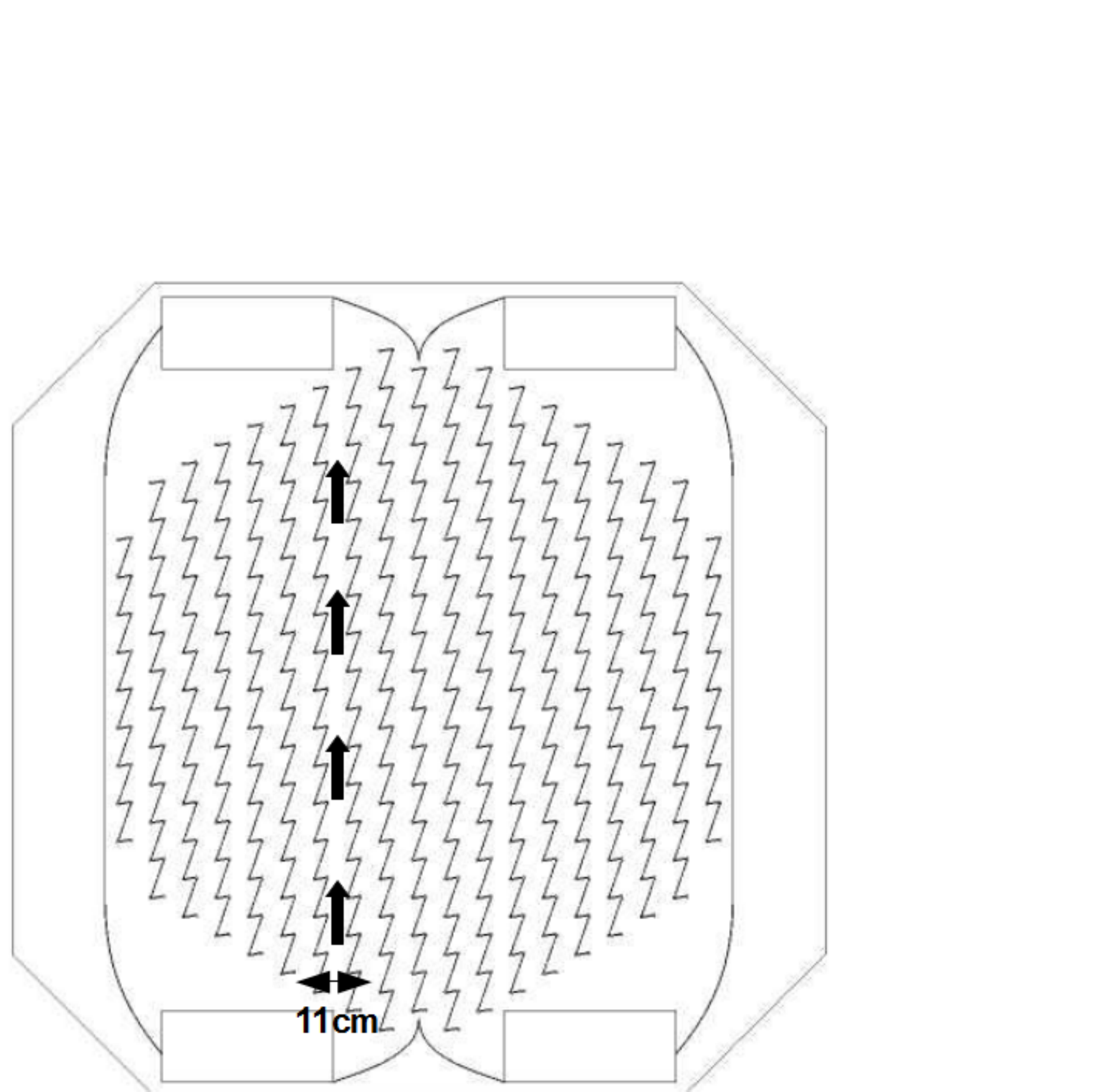}
  \caption{Conceptual drawing showing the top view of the channels for the air circulation. Only walls extending along the optical acisdirection are shown. The rectangles on the top and bottom are ducts to inject and extract the air. The zig-zag line are the readout boards and rails holding them. Finally the curved lines are thin walls confining the circulation in the volume between the plates.}   
  \label{fig:air circulation}
\end{figure}
 
In case of the air cooling system, the empty space between the 35 mm and 5 mm plates is maximized by reducing the auxiliary mechanics in the
clusters. This allows to keep an empty space between two consecutive rows of readout boards that provides channels with a transversal cross section of $11\times 40$ cm$^2$, as shown in Fig. \ref{fig:air circulation}. The air in the camera is forced to pass through a heat exchanger placed in the rear part of the camera by a fancoil. Once cooled down it is injected in the channels formed by consecutive rows of readout boards. The air flow is modulated by ducts from the fancoil to these channels and its speed, whereas the temperature is regulated by the circulation of freon or water in the heat exchanger. The modulation of both the air flow speed and temperature allows to keep the electronics environment temperature in an optimum range and constant in time.   
  
The stiffness of both load bearing designs has been studied by a finite element analysis, resulting in a maximum deformation in the direction of the focal axis below 1 mm for the air cooling design, and of about 2.5 mm for the water cooling one, and negligible in any perpendicular direction for both designs. Studies based in numerical analysis and evaluation of prototypes are being carried out to evaluate the mechanical reliability, maintainability and performance of the associated cooling systems. These will provide the necessary feed back to decide which of the two designs will be used for the LST cameras.

\section{External walls and entrance window}
The external walls of the camera are made with honeycomb panels, which are very
light and strong, white painted to reflect part of the solar radiation during daytime. They are attached to the mechanical tubular structure of the camera using a self-supporting structure. The rear side of the camera is closed by one door which allows access to the rear compartment of the camera for maintenance operations. 

The current design of the entrance window is based on the one of the MAGIC telescopes. It consists of a single sheet of UV transmitting Plexiglas
fixed at the edges of the camera instrumented area by screws. The window is pre-stretched at the moment of assembly to minimize any deformation due 
to its weight. The screws allow to dismount it to access the camera from the front part. Current studies concentrate on verifying the scalability of this design to the LST camera. A challenge is to handle the forces exerted by the interior air on the window, which can reach several thousand Newtons.

\section{Cluster mechanics}\label{sec:cluster}
The cluster mechanics serve two purposes. On the one hand it keeps the relative position of the photodetectors
in a single cluster, and between these and the readout electronics which are rigidly mounted in the cluster.
On the other hand it provides the interface between the cluster and the load bearing structure. Finally, in the
case of the water cooling load bearing structure design, the cluster mechanics should guarantee a good thermal coupling between the electronic components on the readout boards and the cooling plates.

\begin{figure}[t]
  \centering
	\includegraphics[width=0.15\textwidth]{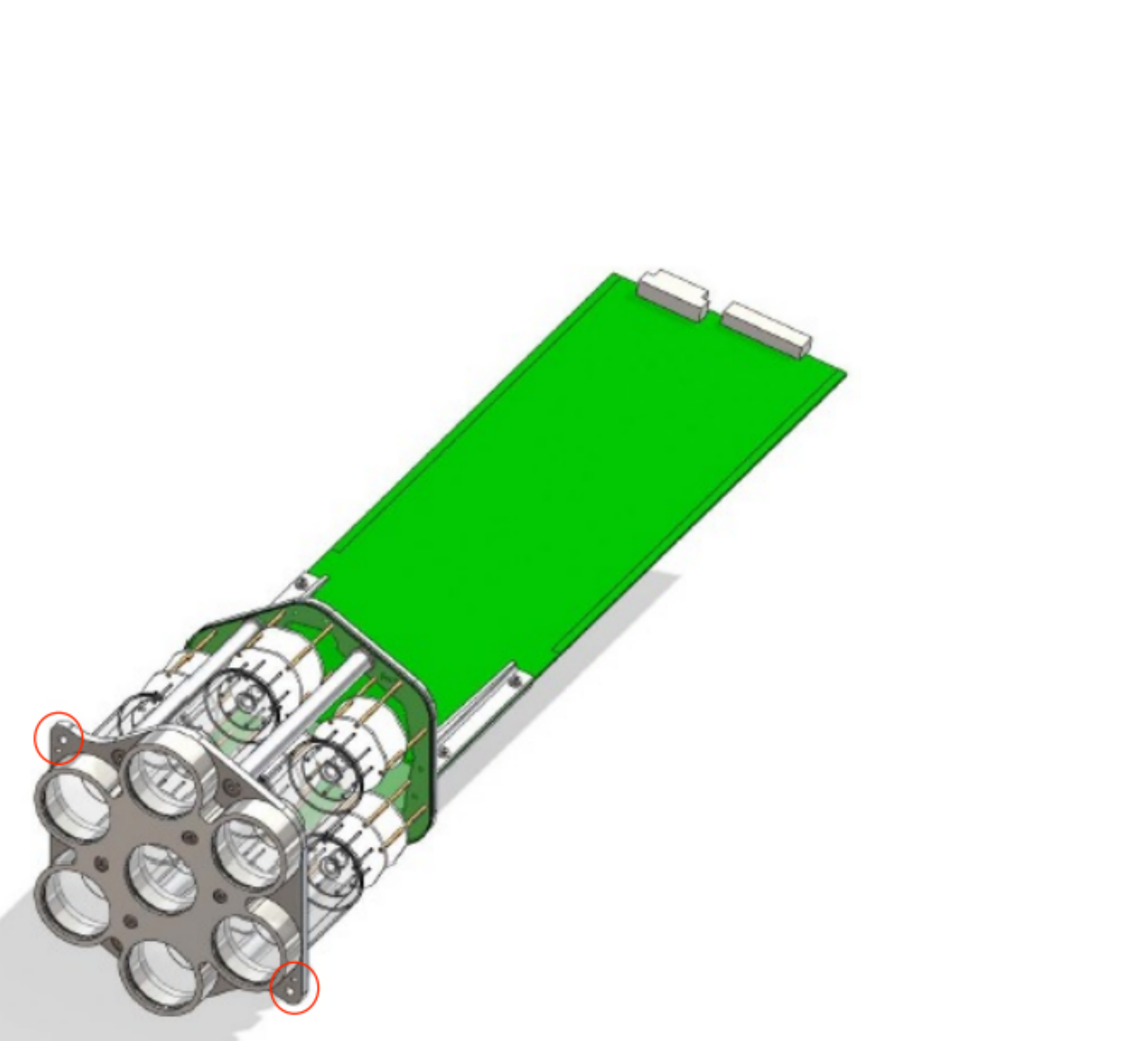}
  \caption{Drawing showing the cluster for the air cooling system. All the mechanic elements are concentrated in the front part. The red circles show the position of the screws fixing the cluster to the load bearing structure.} 
  \label{fig:cluster air cooling}
\end{figure}

In the case of the cluster for the air cooling system, shown in Fig. \ref{fig:cluster air cooling}, the mechanics hold the photomultipliers and front-end electronics, and two clamps on the edges hold the readout electronics. The required elements are mostly made on aluminum with few plastic elements. No further mechanical elements are required since the columns joining the plates provide the remaining support for the electronics. The cluster is mounted on the load bearing mechanics by inserting it from the front, slicing the ensemble through the columns joining the plates and mating the readout board with the backplane bypassing the rear plate with the connectors in the readout board through slits in the former. Finally, two screws fix the front part of the cluster mechanics to the load bearing structure. With this design, all cluster elements remain within the two plates, except of the light guides. 

\begin{figure}[t]
  \centering
	\includegraphics[width=0.4\textwidth]{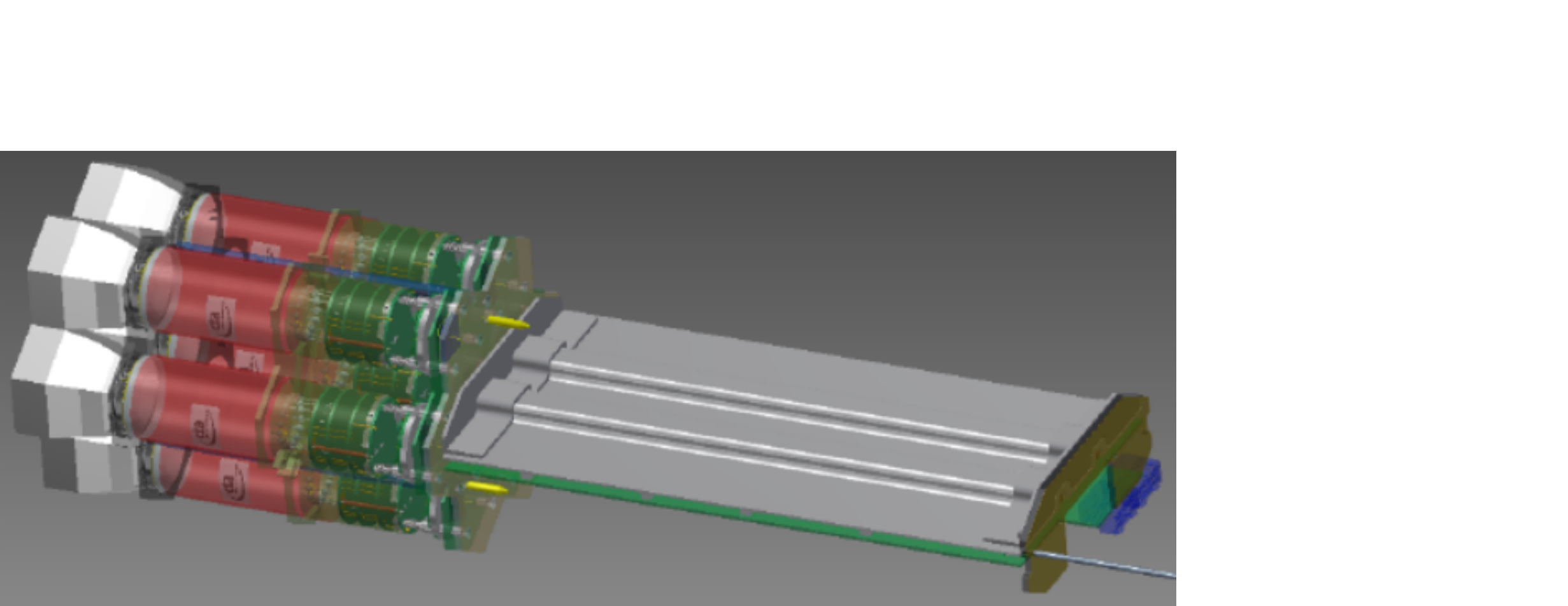}
  \caption{Drawing showing the cluster for the water cooling load bearing structure.}
  \label{fig:cluster water cooling}
\end{figure}

The cluster mechanics designed for the load bearing structure for water cooling is shown in Fig. \ref{fig:cluster water cooling}. Its front part is similar to the air cooling cluster, but provides further elements to guarantee the thermal contact between electronic elements and the load bearing plates. This is achieved by placing an aluminum plate on top of the readout board, in thermal contact with the surface of all the electronics elements. This contact is guaranteed by heat conducting silicon deposited between the readout board and the aluminum plate. The latter is in contact with two heat-pipes that transfer the heat to the edges of the cluster, where it is coupled to
two aluminum angles, which are coated with a  heat conducting rubber on those surfaces which are in contact with the plates. This cluster is inserted from the front of the camera after dismounting the corresponding backplane from the rear part of the camera. Then it is screwed to the rear plate of the load bearing structure and finally it is connected to the back plane. The position of the cluster of the front part is guaranteed by means of the pins shown in yellow in the Fig. \ref{fig:cluster water cooling}, which are inserted in the front plate. With this design only the readout electronics 
remains within the two load bearing structure plates.	
 
\section{Blinds} 

\begin{figure}[t]
  \centering
	\includegraphics[width=0.4\textwidth]{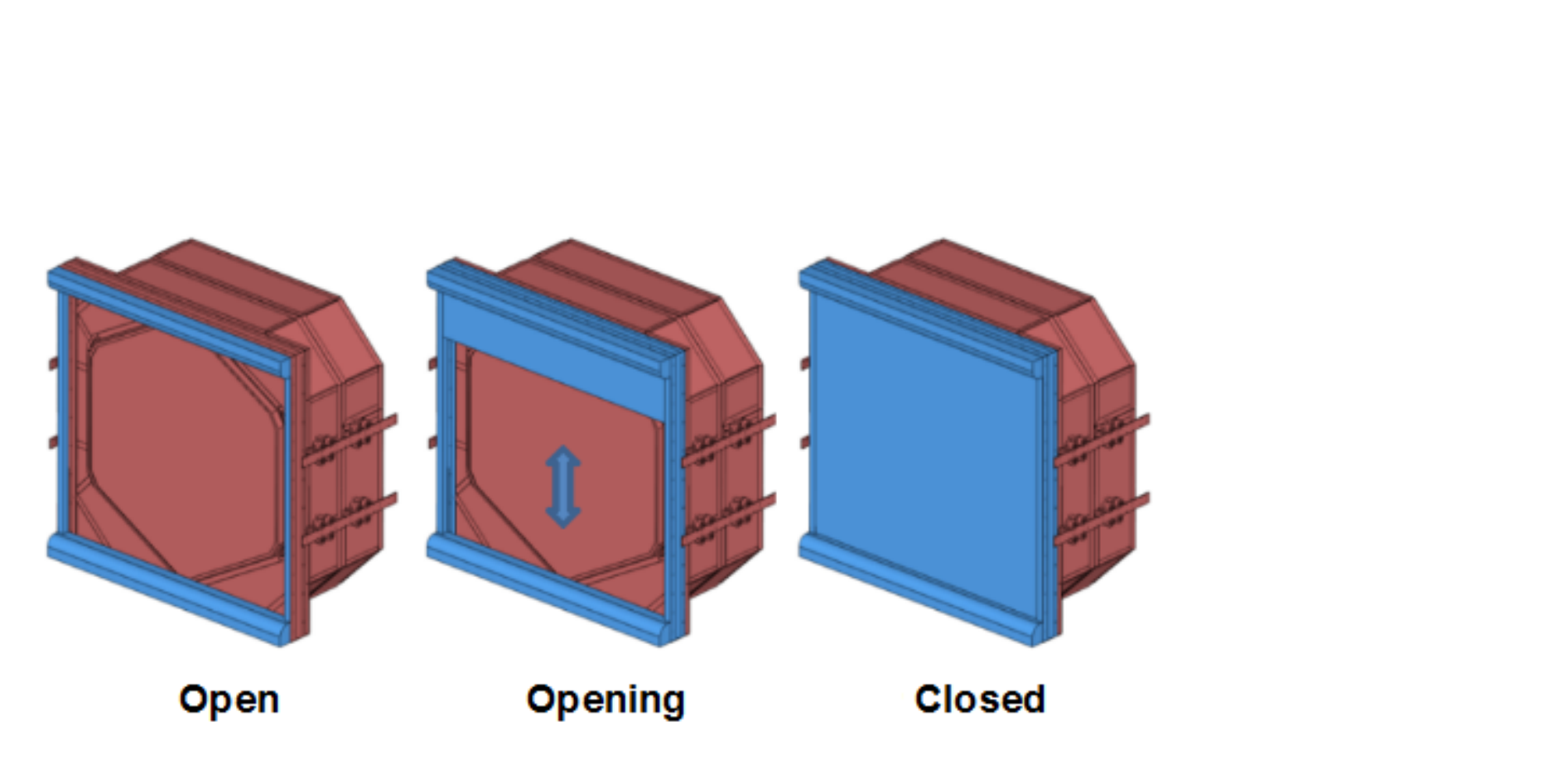}
  \caption{Concept of the blinds system (in blue), showing three stages of its operation.}
  \label{fig:blind concept}
\end{figure}

\begin{figure}[t]
  \centering
	\includegraphics[width=0.3\textwidth]{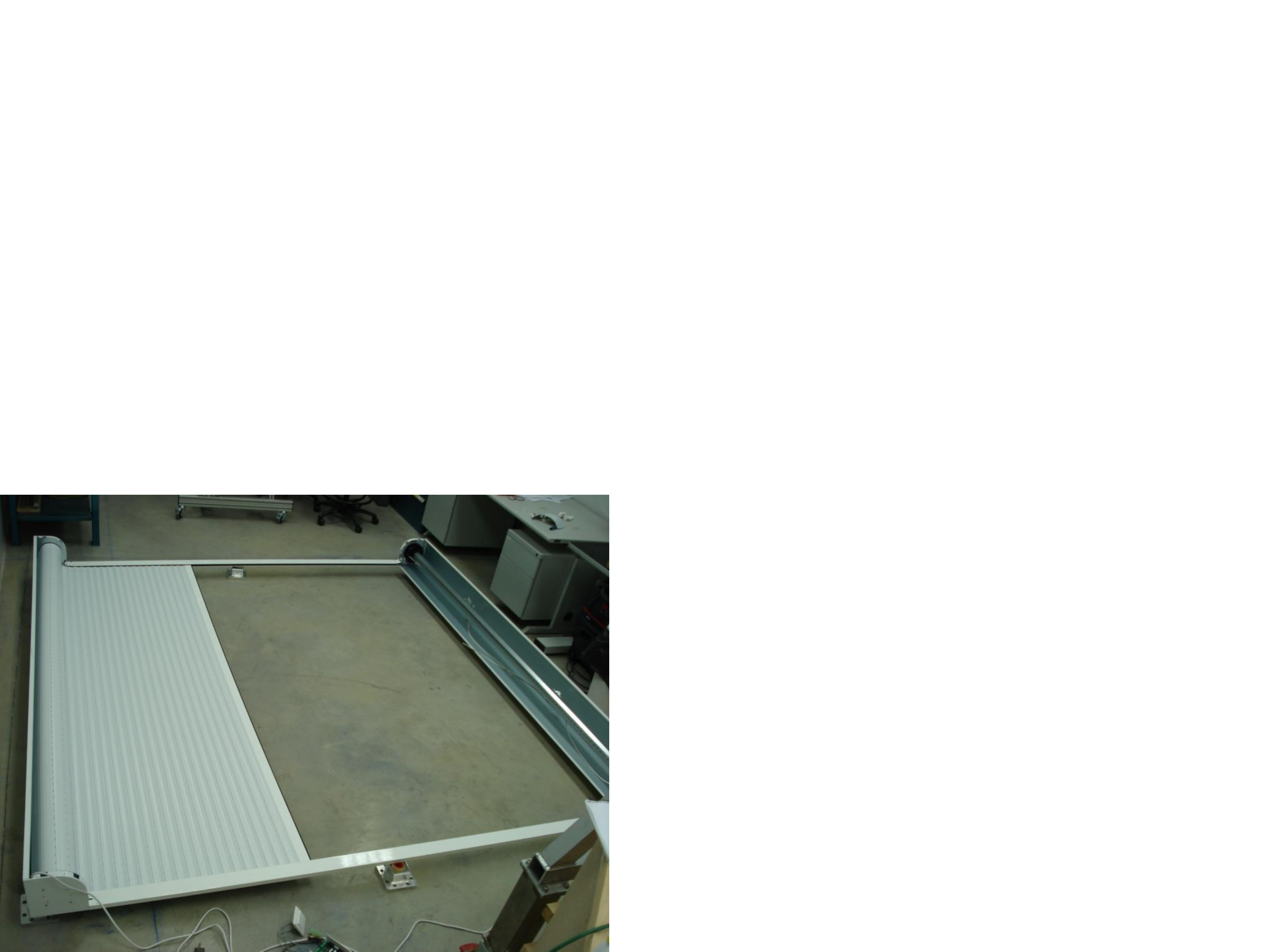}
  \caption{Prototype of the blind system.}
  \label{fig:blind prototype}
\end{figure}

In order to protect the photomultipliers and allow operations during daylight with high 
voltage switched on, a motorized blind completely closes the entrance of light from 
the front of the camera when required. For realiable operation at any camera orientation
it is moved by two motors, one in the lower part of the camera and another of the upper one. The system is mounted
on a frame which is attached by a mechanical interface to the front part of the camera. This
interface keeps a separation between the Plexiglas window that 
protects the photomultipliers and the blind large enough to avoid any interference due
to deformation of the blinds. The blind dimensions are $3\times 3$ m$^2$, and its weight
is about 60 kg. Fig. \ref{fig:blind concept} shows the concept of the system. A prototype of this concept 
has been built and is under study to verify its reliability and mode of operation for any camera
orientation (shown in Fig. \ref{fig:blind prototype}).
\section{Fixation to camera frame}
The fixation of the camera with the camera frame must allow 
displacements of up to 40 cm of the former with respect to the latter 
to compensate possible mis-adjustments of the focal distance. Additionally
it must allow loading and unloading of the camera from the frame for
maintenance operations. The design of this connection is composed of two rails on each side of the camera with two carriages each. The camera-frame connection system is assembled on the structural part of the camera and on dedicated aluminum alloy plates that are part of the camera frame, and the two carriages are assembled to the tubular structure.

\begin{table}[t]
\begin{center}\small
\begin{tabular}{|l|c|c|}
\hline Item & Water cooling& Air cooling \\ \hline
Clusters   &  663 kg & 504 kg \\ \hline
Load bearing structure  & 310 kg  & 275 kg \\ \hline
Tubular  structure & 250 kg  & 250 kg \\ \hline
Fans and ducts &  0 kg & 80 kg \\ \hline
Water and pipes & 20 kg & 0 kg \\ \hline
Total & 1243 kg & 1109 kg \\ \hline
\end{tabular}
\caption{Weight of main components of the LST camera.}
\label{tab:weights}
\end{center}
\end{table}  

\section{Conclusions}
Given these cluster designs and the ones for the load bearing structure, it is possible to estimate the weight of the camera, shown in table \ref{tab:weights}. This table shows the weight contraint of 2000 kg for the whole camera is fullfiled if the weight of the remaining camera elements is less than 750 kg. Final details of the design have to be fixed once the optimum cooling strategy is decided based on tests being carried out. 

\section{Acknowledgements}
We gratefully acknowledge support from the agencies and organizations listed in this page: \emph{http://ww.cta-observatory.org/?q=node/22}

\end{document}